\PassOptionsToPackage{unicode}{hyperref}
\PassOptionsToPackage{hyphens}{url}
\documentclass[
  11pt,
]{article}

\usepackage{tikz}
\usetikzlibrary{positioning, shapes.geometric, arrows.meta}
\usepackage{xcolor}
\usepackage{graphicx}

\definecolor{customblue}{RGB}{26, 115, 232}
\definecolor{bgblue}{RGB}{240, 244, 255}
\definecolor{bordergray}{RGB}{180, 180, 180}

\usepackage{amsmath,amssymb}
\usepackage{iftex}
\ifPDFTeX
  \usepackage[T1]{fontenc}
  \usepackage[utf8]{inputenc}
  \usepackage{textcomp} 
\else 
  \usepackage{unicode-math} 
  \defaultfontfeatures{Scale=MatchLowercase}
  \defaultfontfeatures[\rmfamily]{Ligatures=TeX,Scale=1}
\fi
\ifPDFTeX
  \IfFileExists{lmodern.sty}{\usepackage{lmodern}}{}
\else
  \usepackage{fontspec}
\fi
\ifPDFTeX\else
\fi
\IfFileExists{upquote.sty}{\usepackage{upquote}}{}
\IfFileExists{microtype.sty}{
  \usepackage[]{microtype}
  \UseMicrotypeSet[protrusion]{basicmath} 
}{}
\makeatletter
\@ifundefined{KOMAClassName}{
  \IfFileExists{parskip.sty}{%
    \usepackage{parskip}
  }{
    \setlength{\parindent}{0pt}
    \setlength{\parskip}{6pt plus 2pt minus 1pt}}
}{
  \KOMAoptions{parskip=half}}
\makeatother
\usepackage{xcolor}
\usepackage[margin=1in]{geometry}
\usepackage{longtable,booktabs,array}
\usepackage{calc} 
\usepackage{etoolbox}
\makeatletter
\patchcmd\longtable{\par}{\if@noskipsec\mbox{}\fi\par}{}{}
\makeatother
\IfFileExists{footnotehyper.sty}{\usepackage{footnotehyper}}{\usepackage{footnote}}
\makesavenoteenv{longtable}
\providecommand{\tightlist}{%
  \setlength{\itemsep}{0pt}\setlength{\parskip}{0pt}}
\ifLuaTeX
  \usepackage{selnolig}  
\fi
\IfFileExists{bookmark.sty}{\usepackage{bookmark}}{\usepackage{hyperref}}
\IfFileExists{xurl.sty}{\usepackage{xurl}}{} 
\hypersetup{
  pdftitle={An Open-Source Data Infrastructure for Alloy Synthesis Research: Stage 1 -\/-\/- Composition Screening and Literature Cross-Referencing},
  pdfauthor={Iliya Radulov},
  hidelinks,
  pdfcreator={LaTeX via pandoc}}

\title{An Open-Source Data Infrastructure for Alloy Synthesis Research: Stage 1 -\/-\/- Composition Screening and Literature Cross-Referencing}
\author{Iliya Radulov}
\date{October 2026}

\begin{document}
\maketitle

\begin{abstract}
Experimental materials-science labs routinely accumulate synthesis and characterization data across spreadsheets, instrument export files, and ad-hoc folders, with no durable link between a sample's recipe, its measured properties, and whether a similar composition has already been reported elsewhere. We present the first stage of an open-source data infrastructure built to close this loop for a single working alloy-synthesis lab: a relational schema for composition and synthesis records, a strict composition/mass calculator, three empirical alloy-design screening metrics (VEC, atomic-size mismatch, mixing enthalpy) together with a melt/boil-point synthesis-feasibility check, an automated cross-referencing layer against three public computational-materials databases (Materials Project, OQMD, Alexandria) with deduplication and adjustable relevance filtering, and a desktop GUI that ties the whole workflow to a single ``new entry'' action. The system is built and used by a practicing researcher rather than a software team, and we report it as a systems/methodology contribution: the architecture, the concrete engineering failures encountered while building it, and how each was diagnosed and fixed, rather than new materials-science results. We argue the specific defects found --- a parser ambiguity that produced silent wrong answers, a partial code edit that broke a dependency chain invisibly, a database-metadata bug traceable to an un-cast numeric type, and a credential-hygiene gap that \texttt{.gitignore} alone did not close --- generalize beyond this codebase to any small scientific-software project built by a domain expert without a software-engineering background. The system and the lessons from building it are reported here; two follow-on papers cover the validated characterization pipelines (Stage 2) and the retrieval-augmented synthesis advisor and CALPHAD thermodynamics module (Stage 3) built on top of this foundation.
\end{abstract}

\hypertarget{introduction}{%
\section{Introduction}\label{introduction}}

A working experimental alloy lab generates data at several disconnected points: a hand-written synthesis recipe, a diffractometer's raw output file, a magnetometer's hysteresis loop, a micrograph, and a researcher's memory of whether a similar composition has been tried before. In most small labs --- including the one this system was built for --- these pieces live in spreadsheets, instrument-vendor file formats, and loosely named folders, linked only by the researcher's own recall. This is adequate for a single project but breaks down precisely when it would be most valuable: when deciding whether a new composition is worth attempting, and when later trying to assemble a dataset for machine learning.

This paper reports the first stage of a response to that problem: an open-source data infrastructure built by a single materials-science researcher, without a software-engineering background, over several months of part-time development. Stage 1, reported here, establishes the core loop --- composition in, screening metrics and literature cross-checks out, everything persisted to one queryable store --- and is deliberately scoped to stop there. Two later stages, each reported separately, extend it: Stage 2 adds validated pipelines that extract quantitative results from raw instrument files (XRD, VSM, SEM) and attach them to the same sample record, and Stage 3 adds a retrieval-augmented literature advisor and a CALPHAD-based thermodynamic feasibility check.

We deliberately frame this as a software-systems contribution rather than a materials-science one. The empirical screening metrics used (Section~\ref{composition-screening}) are established in the high-entropy-alloy literature; the contribution is the infrastructure that makes them usable as a daily research habit rather than a one-off calculation, and the concrete, reproducible account of what went wrong while building that infrastructure and why. We believe this account is useful specifically because it comes from a domain scientist building research software without formal training in it --- a situation common across experimental science, and one the software-engineering literature rarely documents from the inside.

\textbf{Contributions.}

\begin{enumerate}
\def\labelenumi{\arabic{enumi}.}
\tightlist
\item
  An open-source, end-to-end data-capture architecture for alloy synthesis research, combining a composition calculator, physics-based screening metrics, and automated multi-database literature cross-referencing behind a single desktop workflow.
\item
  A deduplication and adjustable-relevance-cutoff scheme for reconciling literature-search results across three structurally different public materials databases.
\item
  A documented, reproducible account of five engineering findings --- four defects in our own code (a silent parsing ambiguity, a dependency-breaking partial edit, a numeric-type database bug, and a credential-hygiene gap) and one dependency-level rejection (a third-party database client discarded after it was found to corrupt results silently) --- each generalized into a lesson for similarly-resourced scientific software projects (Section~\ref{engineering-validation-and-lessons-learned}).
\item
  A scoping discussion of what Stage 1 deliberately excludes, and how it hands off to the Stage 2 and Stage 3 systems built on top of it (Sections~\ref{discussion-and-limitations}--\ref{conclusion-and-future-work}).
\end{enumerate}

The rest of this paper is organized as follows. Section~\ref{related-work-and-positioning} positions the system relative to electronic lab notebooks and public materials-data repositories. Section~\ref{system-overview} gives a system-level overview. Section~\ref{architecture-and-implementation} describes the architecture and implementation in detail. Section~\ref{engineering-validation-and-lessons-learned} reports the engineering defects found and the generalizable lessons drawn from them. Section~\ref{discussion-and-limitations} discusses scope and limitations, and Section~\ref{conclusion-and-future-work} concludes.

\hypertarget{related-work-and-positioning}{%
\section{Related Work and Positioning}\label{related-work-and-positioning}}

\textbf{Electronic lab notebooks (ELNs).} General-purpose ELN platforms provide structured note-taking, file attachment, and search, but treat composition and screening as free text or custom fields rather than first-class, queryable, computed data. They do not natively compute VEC/delta/mixing-enthalpy from a formula string, nor cross-reference a new entry against external materials databases at entry time. The system described here can be read as a narrow, deeply specialized alternative to a general ELN for exactly one recurring task --- ``is this alloy composition already known, and is it synthesizable'' --- at the cost of being far less general.

\textbf{Public computational-materials databases.} Materials Project, OQMD, and Alexandria (Section~\ref{multi-database-literature-cross-referencing}) are themselves large, well-engineered systems for discovering and querying computed materials properties, each with its own API and client library. They are built for materials discovery --- finding new candidate materials with target properties --- not for a practicing synthesis lab's narrower question of whether a specific composition already under consideration is known and how it might be made. This system does not reimplement or compete with them; it is a thin, opinionated orchestration layer that queries all three automatically for every new sample, normalizes their differing result shapes into one comparable structure, and removes the manual, error-prone, multi-tab-browser workflow that querying them individually would otherwise require.

\textbf{High-entropy/multi-principal-element alloy design metrics.} The VEC, $\delta$, and $\Delta H_{\text{mix}}$ screening criteria implemented here (Section~\ref{composition-screening}) are established empirical rules from the high-entropy-alloy design literature, used to estimate phase-formation tendency from composition alone before any synthesis attempt. Our contribution is not a new metric but making these metrics a zero-friction, always-computed part of every new sample entry, with full periodic-table coverage and an explicit failure mode (rather than a silently wrong number) when element data is incomplete.

\textbf{Positioning.} This system sits between a general ELN and a materials-discovery database: narrower than either, but built to make one specific, recurring researcher decision --- should I attempt to synthesize this composition, and how --- fast, consistent, and automatically cross-checked, in a tool scaled to a single lab rather than an institution.

\hypertarget{system-overview}{%
\section{System Overview}\label{system-overview}}

\textbf{Infrastructure evolution and a recommended starting setup.} The infrastructure behind this system was not designed up front; it grew iteratively, in step with the project's own requirements, through four distinct configurations: (1) a single laptop running the application, screening logic, and a local PostgreSQL instance together; (2) a more capable laptop, adopted once composition screening and multi-database literature queries made the original machine a bottleneck; (3) a split configuration, with the canonical database moved to an always-on home-network server while the laptop remained the interactive client, adopted once the database needed to be reachable consistently rather than only while the development laptop happened to be open; and (4) the current three-role configuration, in which a dedicated GPU-equipped desktop performs the heavier downstream computation introduced in later stages (local large-language-model inference for the Stage 3 RAG advisor, CALPHAD equilibrium calculations), the home server remains the sole canonical store for both the database and raw data files, and the original laptop now serves purely as a lightweight access point into the rest of the system.

We report this evolution rather than only a final architecture because we believe the lesson is in the trajectory, not just the endpoint. A deployment that intends to stay within Stage 1's scope --- composition entry, screening, and literature cross-referencing for a single researcher --- is well served by a single machine with a local PostgreSQL instance, and we do not recommend setting up distributed infrastructure prematurely. A deployment that anticipates growing toward the shared-access, characterization-pipeline, or compute-heavy components of later stages, however, is better served by establishing the three-role split --- a thin access/development client, an always-on server as the single canonical data store, and a separate compute-capable machine for heavy processing --- from the outset, rather than migrating mid-project as we did twice. Database layout (Section~\ref{database-schema}) and the server migration that made the shared-access role possible are reported in full in Stage 2.

\textbf{Design goals.} The system was designed around three goals, derived directly from the lab's prior pain points: (1) every new composition should be screened and cross-checked against prior art automatically, as a side effect of normal data entry rather than a separate manual step; (2) every record, once entered, should be permanently and queryably linked to its full context --- synthesis route, screening result, literature-check outcome --- rather than living as an isolated spreadsheet row; (3) the system should degrade gracefully on incomplete or ambiguous input (missing element data, ambiguous formula strings, unreachable network services) by failing loudly and locally rather than silently producing a wrong answer or crashing.

\textbf{Scope boundary.} Stage 1, as reported here, ends at composition, screening, and literature cross-referencing. It accepts and stores raw characterization instrument files (XRD/VSM/SEM) for visualization and record-keeping --- including sorting them by file type and sample ID (Section~\ref{desktop-gui-and-data-sorting-utility}) --- but does not yet extract quantitative results from them; that extraction pipeline, built and separately validated afterward, is Stage 2. It also does not yet include the retrieval-augmented synthesis advisor or thermodynamic phase-diagram tooling of Stage 3. This scoping was a conscious choice: each stage was brought to a stable, validated state before the next was started, rather than developing all three in parallel.

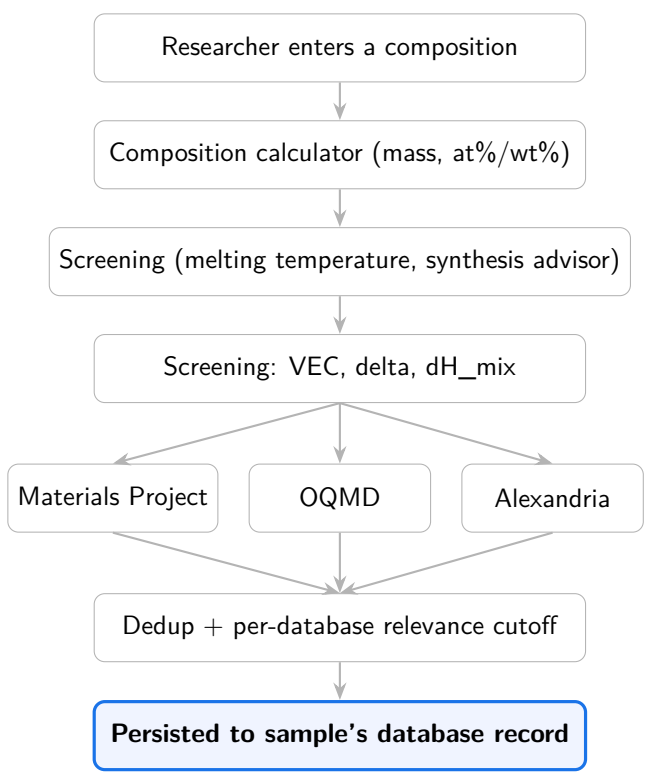
\begin{figure}[htbp]
\centering
\begin{tikzpicture}[
    node distance = 0.5cm and 0.4cm,
    block/.style = {
        rectangle, 
        rounded corners=4pt, 
        draw=bordergray, 
        minimum width=6.5cm, 
        minimum height=0.9cm, 
        align=center,
        font=\sffamily\small
    },
    smallblock/.style = {
        rectangle, 
        rounded corners=4pt, 
        draw=bordergray, 
        minimum width=2.4cm, 
        minimum height=0.9cm, 
        align=center,
        font=\sffamily\small
    },
    blueblock/.style = {
        rectangle, 
        rounded corners=4pt, 
        draw=customblue, 
        fill=bgblue,
        line width=1.2pt,
        minimum width=6.5cm, 
        minimum height=0.9cm, 
        align=center,
        font=\sffamily\small\bfseries
    },
    arrow/.style = {
        ->, 
        >=Stealth, 
        draw=bordergray, 
        thick
    }
]

    \node (n1) [block] {Researcher enters a composition};
    \node (n2) [block, below=of n1] {Composition calculator (mass, at\%/wt\%)};
    \node (n3) [block, below=of n2] {Screening (melting temperature, synthesis advisor)};
    \node (n4) [block, below=of n3] {Screening: VEC, delta, dH\_mix};
    
    \node (n5b) [smallblock, below=0.8cm of n4] {OQMD};
    \node (n5a) [smallblock, left=of n5b] {Materials Project};
    \node (n5c) [smallblock, right=of n5b] {Alexandria};
    
    \node (n6) [block, below=0.8cm of n5b] {Dedup + per-database relevance cutoff};
    \node (n7) [blueblock, below=of n6] {Persisted to sample's database record};

    \draw [arrow] (n1) -- (n2);
    \draw [arrow] (n2) -- (n3);
    \draw [arrow] (n3) -- (n4);
    
    \draw [arrow] (n4.south) -- (n5a.north);
    \draw [arrow] (n4.south) -- (n5b.north);
    \draw [arrow] (n4.south) -- (n5c.north);
    
    \draw [arrow] (n5a.south) -- (n6.north);
    \draw [arrow] (n5b.south) -- (n6.north);
    \draw [arrow] (n5c.south) -- (n6.north);
    
    \draw [arrow] (n6) -- (n7);

    \node [above=0.6cm of n1.north west, anchor=south west, align=left, font=\sffamily] {
        \textbf{\large One entry triggers screening and three database cross-checks automatically} \\
        \textcolor{gray!90!black}{\footnotesize Stage 1 end-to-end workflow}
    };

\end{tikzpicture}
\caption{All three literature databases are queried in parallel for every new entry; results are then deduplicated and filtered before anything is written to the database, so the record a researcher sees already reflects the current relevance cutoff rather than a raw, unfiltered dump.}
\label{fig:workflow}
\end{figure}

Figure~\ref{fig:workflow} shows the Stage 1 workflow end to end, from a researcher entering a composition to a persisted, cross-checked database record.

\hypertarget{architecture-and-implementation}{%
\section{Architecture and Implementation}\label{architecture-and-implementation}}

\hypertarget{database-schema}{%
\subsection{Database schema}\label{database-schema}}

The system persists to a PostgreSQL database with a \texttt{samples} table as its hub, holding composition (JSONB, for schema flexibility) alongside dedicated numeric columns for the three screening metrics so they remain directly indexable and filterable. Related tables cover \texttt{synthesis}, \texttt{characterization}, \texttt{properties}, \texttt{compositions} (normalized element fractions), \texttt{literature\_sources}, \texttt{material\_classes}, and \texttt{literature\_checks} (one row per deduplicated candidate phase per source database per sample). Credentials are isolated in a \texttt{.env} file read via \texttt{python-dotenv}; the connection layer fails loudly rather than silently defaulting when a required variable is absent.

\begin{figure}[htbp]
\centering
\begin{tikzpicture}[
    node distance = 0.9cm and 0.4cm,
    tbl/.style = {
        rectangle,
        rounded corners=3pt,
        draw=bordergray,
        minimum width=2.7cm,
        minimum height=1.0cm,
        align=center,
        font=\sffamily\scriptsize
    },
    hubtbl/.style = {
        rectangle,
        rounded corners=3pt,
        draw=customblue,
        fill=bgblue,
        line width=1.2pt,
        minimum width=2.9cm,
        minimum height=1.0cm,
        align=center,
        font=\sffamily\scriptsize\bfseries
    },
    looseTbl/.style = {
        rectangle,
        rounded corners=3pt,
        draw=bordergray,
        dashed,
        minimum width=2.9cm,
        minimum height=1.1cm,
        align=center,
        font=\sffamily\scriptsize
    },
    fkarrow/.style = {
        ->,
        >=Stealth,
        draw=bordergray,
        thick
    },
    fklabel/.style = {
        font=\sffamily\tiny,
        fill=white,
        inner sep=1pt
    }
]

\node (samples) [hubtbl] {\textbf{samples}\\[1pt] \mdseries id (PK)};

\node (matclass) [tbl, above=1.3cm of samples] {material\_classes\\[1pt] id (PK)};

\node (synthesis) [tbl, below=1.7cm of samples, xshift=-4.65cm] {synthesis\\[1pt] id (PK)\\ sample\_id (FK)};
\node (compositions) [tbl, right=0.4cm of synthesis] {compositions\\[1pt] id (PK)\\ sample\_id (FK)};
\node (charac) [tbl, right=0.4cm of compositions] {characterization\\[1pt] id (PK)\\ sample\_id (FK)};
\node (litchecks) [tbl, right=0.4cm of charac] {literature\_checks\\[1pt] id (PK)\\ sample\_id (FK)};

\node (properties) [tbl, below=1.2cm of charac] {properties\\[1pt] id (PK)\\ characterization\_id (FK)};

\node (litsources) [looseTbl, above=1.3cm of litchecks] {literature\_sources\\[1pt] id (PK)};

\draw [fkarrow] (samples.north) -- (matclass.south) node[fklabel, midway, right] {material\_class\_id};
\draw [fkarrow] (synthesis.north) -- ++(0,0.25) node[fklabel, above] {sample\_id} -| (samples.south);
\draw [fkarrow] (compositions) -- (samples) node[fklabel, midway, right] {sample\_id};
\draw [fkarrow] (charac.north) -- ++(0,0.25) node[fklabel, above] {sample\_id} -| (samples.south);
\draw [fkarrow] (litchecks.north) -- ++(0,0.25) node[fklabel, above] {sample\_id} -| (samples.south);
\draw [fkarrow] (properties) -- (charac) node[fklabel, midway, right] {characterization\_id};
\draw [fkarrow] (samples.east) to [out=20,in=70,looseness=6] node[fklabel, above right] {parent\_sample\_id} (samples.north east);

\end{tikzpicture}
\caption{Core Stage 1 schema. \texttt{samples} is the hub table; \texttt{synthesis}, \texttt{compositions}, \texttt{characterization}, and \texttt{literature\_checks} each reference it by \texttt{sample\_id}, and \texttt{samples.parent\_sample\_id} self-references for derived/related samples. \texttt{properties} attaches to \texttt{characterization}. \texttt{literature\_sources} (dashed) is populated with source DOIs from Materials Project's text-mined synthesis records when a sample is saved, but is not currently linked by a foreign key from \texttt{literature\_checks}.}
\label{fig:schema}
\end{figure}

Figure~\ref{fig:schema} shows the core schema just described: \texttt{samples} as the hub, the four tables that reference it by \texttt{sample\_id}, the self-referencing \texttt{parent\_sample\_id}, and \texttt{properties} attaching to \texttt{characterization}.

\hypertarget{composition-calculator}{%
\subsection{Composition calculator}\label{composition-calculator}}

The calculator parses a composition string, converts between atomic and weight percent, and computes synthesis masses. Its formula parser is deliberately strict: every element must be followed by an explicit numeric fraction (e.g.\ \texttt{Nd2Fe14B1}, never \texttt{Nd2Fe14B}). This was not the original design --- an earlier, permissive parser treated a genuinely ambiguous input like ``Cp'' (which could mean the elements C and P, or an unrecognized token) inconsistently rather than flagging it, a defect discussed in detail in Section~\ref{silent-parsing-ambiguity}. The calculator additionally supports two modes reflecting real synthesis practice: pre-alloy-based synthesis (starting from a commercial pre-alloyed ingot of known composition) and evaporation-loss compensation (an excess-fraction term for elements known to evaporate during melting), including an inverse mode that, given a fixed mass of pre-alloy on hand, solves for the raw element masses needed to reach a target composition while correctly separating contributions from elements that appear in both the pre-alloy and the raw additions.

\hypertarget{composition-screening}{%
\subsection{Composition screening}\label{composition-screening}}

Three metrics are computed for every composition: valence electron concentration (VEC) {[}1{]}, atomic-size mismatch ($\delta$) {[}2{]}, and mixing enthalpy ($\Delta H_{\text{mix}}$) {[}3,4{]}, established empirical indicators of phase-formation tendency in multi-principal-element alloy design. VEC is computed as the composition-weighted mean of elemental valence electron counts; $\delta$ as the root-mean-square deviation of atomic radii from the composition-weighted mean; $\Delta H_{\text{mix}}$ via a pairwise Miedema-style parameter table {[}3,4{]}. VEC and $\delta$ draw on per-element physical constants spanning the full periodic table (103 elements). These metrics flag phase-formation tendency from composition alone; Section~\ref{multi-database-literature-cross-referencing} narrows that estimate using existing literature, and the CALPHAD-based module reported in the Stage 3 paper narrows it further for compositions with a partially known chemical subsystem.

\hypertarget{silent-data-table-defect}{%
\subsubsection{A silent data-table defect in \texorpdfstring{$\Delta H_{\text{mix}}$}{ΔH\_mix}}\label{silent-data-table-defect}}

$\Delta H_{\text{mix}}$ originally used a pairwise mixing-enthalpy table whose literature source was not recorded at implementation time, and cross-checking all 40 original pairs against Takeuchi and Inoue's tabulated Miedema-model values {[}3{]} (confirmed internally consistent with their earlier compilation of the same binary systems {[}4{]}) surfaced two independent problems. First, four pairs differed in sign --- including the Fe--Nd pair, whose original value was opposite in sign to the tabulated one --- a defect that would have silently stabilised compositions where the literature indicates destabilisation. Second, the regular-solution formula underlying these tabulated values,
\[
\Delta H^{\text{chem}} = \Sigma \, \Omega_{ij}\, c_i\, c_j, \quad \text{with } \Omega_{ij} = 4 \times \Delta H_{AB}^{\text{mix}} \ \text{{[}4{]}},
\]
requires each binary value to be scaled by a factor of four before use as a pairwise interaction parameter, a scaling the original implementation omitted; the implementation also defaulted silently to zero for any pair absent from the table, masking missing coverage as a weak-interaction result --- the same failure mode as the parser defect of Section~\ref{silent-parsing-ambiguity}, one layer deeper in the data pipeline.

The table and formula have since been replaced. $\Delta H_{\text{mix}}$ is now stored as the literature $\Delta H_{AB}^{\text{mix}}$ value for each pair, expanded well beyond the original 40; all entries are from Takeuchi \& Inoue 2005 {[}3{]}, cross-checked entry by entry against a machine-readable copy of that table, and no entries remain flagged. The $\times 4$ regular-solution factor is applied explicitly inside the calculation and verified by a self-test confirming that an equiatomic binary reduces to exactly $\Delta H_{AB}^{\text{mix}}$. An untabulated pair now raises an explicit error rather than defaulting to zero.

\hypertarget{incomplete-data-failure-modes}{%
\subsubsection{Incomplete-data failure modes}\label{incomplete-data-failure-modes}}

Two exceptions enforce the same design principle --- a failure is loud and local, never a plausible-looking wrong number. \texttt{IncompleteElementDataError} is raised whenever a composition includes an element outside the tabulated element data; the newly added \texttt{IncompletePairDataError} is raised analogously whenever two tabulated elements lack a recorded $\Delta H_{\text{mix}}$ pair. Both the CLI tool and the GUI catch these and degrade gracefully (screening is skipped for that run, with the affected columns left NULL) rather than either crashing or persisting a value that looks valid but was computed from an incomplete element or pair set. This is the same principle as the strict formula parser (Section~\ref{composition-calculator}) and the connection layer's refusal to silently default when a required environment variable is absent (Section~\ref{database-schema}): the only safe response to incomplete input is to refuse to compute at all.

\hypertarget{synthesis-feasibility-check}{%
\subsubsection{Synthesis-feasibility check}\label{synthesis-feasibility-check}}

A related but distinct check runs alongside the three metrics. \texttt{check\_synthesis\_feasibility} compares the melting and boiling points of the constituent elements, rather than estimating phase-formation tendency from composition. It finds the highest melting point and the lowest boiling point among the elements present and reports the margin between them, classifying the result as ok, caution, blocked, or unknown (when melt/boil data for an element is missing). A blocked composition --- where an element would boil before another has fully melted under straightforward arc or induction melting --- returns a list of suggested alternative routes (mechanical alloying by ball milling, powder sintering, diffusion bonding); a caution composition returns softer guidance (monitor mass loss during melting, check for literature precedent, run a small-scale test melt first).

Unlike VEC, $\delta$, and $\Delta H_{\text{mix}}$, this is not a phase-formation indicator from the high-entropy-alloy literature. It is a practical feasibility check on the synthesis route itself, reported here because it is the one screening output whose result is directly actionable --- a blocked verdict names a specific alternative route, whereas the phase-formation metrics only narrow the space of compositions worth attempting. The distinction is deliberate: the three metrics answer ``what phases might form,'' the feasibility check answers ``can this be made by the route I have.'' The two together are what makes a new composition entry useful before any experimental work begins.

\begin{figure}[htbp]
    \centering
    \includegraphics[width=1\textwidth]{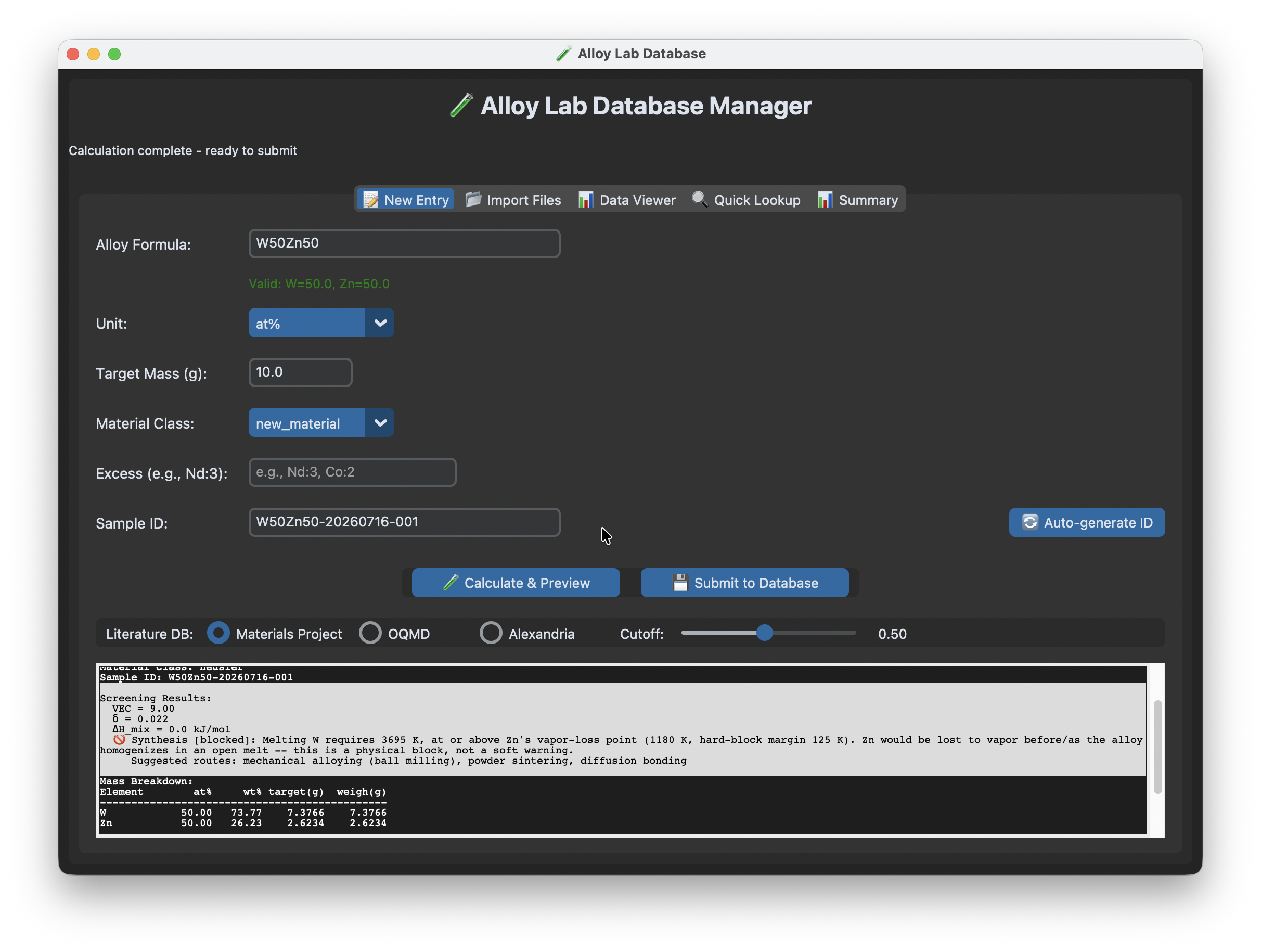}
    \caption{The New Entry tab after ``Calculate \& Preview'': composition parsing, mass breakdown, the three screening metrics, and a blocked synthesis-feasibility verdict with suggested alternative routes, all from a single action.}
    \label{fig:new-entry-gui}
\end{figure}

Figure~\ref{fig:new-entry-gui} shows this in practice, on the New Entry tab: entering a composition (W\textsubscript{50}Zn\textsubscript{50}) triggers the composition calculator, the three screening metrics, and the feasibility check in one action, returning here a blocked verdict --- Zn would be lost to vapor before W melts --- together with the suggested alternative synthesis routes.

\hypertarget{multi-database-literature-cross-referencing}{%
\subsection{Multi-database literature cross-referencing}\label{multi-database-literature-cross-referencing}}

Every new composition is checked against three public materials databases before being logged:

\begin{longtable}[]{@{}
  >{\raggedright\arraybackslash}p{(\columnwidth - 4\tabcolsep) * \real{0.3333}}
  >{\raggedright\arraybackslash}p{(\columnwidth - 4\tabcolsep) * \real{0.3333}}
  >{\raggedright\arraybackslash}p{(\columnwidth - 4\tabcolsep) * \real{0.3333}}@{}}
\toprule\noalign{}
\begin{minipage}[b]{\linewidth}\raggedright
Database
\end{minipage} & \begin{minipage}[b]{\linewidth}\raggedright
Access
\end{minipage} & \begin{minipage}[b]{\linewidth}\raggedright
Distinguishing data
\end{minipage} \\
\midrule\noalign{}
\endhead
\bottomrule\noalign{}
\endlastfoot
Materials Project & Official \texttt{mp-api} client & DFT hull distance, text-mined synthesis recipes; oxide-heavy coverage \\
OQMD & OPTIMADE protocol & \texttt{experimentally\_known} flag; no API key required \\
Alexandria & OPTIMADE protocol & Much larger (2.5M+ entries) computational dataset; no experimental flag \\
\end{longtable}

A fourth candidate, AFLOW, was evaluated and ultimately rejected for reasons that go beyond ordinary integration friction; see Section~\ref{a-dependency-rejected-for-silent-corruption-risk} for the full account of why it was dropped and what that implies more generally about adopting unmaintained scientific-software dependencies.

\hypertarget{deduplication-and-relevance-filtering}{%
\subsubsection{Deduplication and relevance filtering}\label{deduplication-and-relevance-filtering}}

The three working sources are normalized through a shared adapter layer (see Section~\ref{desktop-gui-and-data-sorting-utility} for how results are then surfaced to the researcher) into one common candidate shape, then processed by two stages: deduplication by formula (preferring experimentally-known results, then most stable, then closest composition match), and a per-database, user-adjustable distance cutoff --- added after real use showed Alexandria's much larger dataset returning dozens of only loosely related formulas per query, more volume than deduplication alone resolves. Both the OQMD and Alexandria clients carry custom retry logic for transient server errors, since the underlying OPTIMADE client library's built-in retry only triggers on HTTP 429 and otherwise surfaces a failure without retrying.

\hypertarget{four-tier-synthesis-feasibility-classification}{%
\subsubsection{Four-tier literature-match classification}\label{four-tier-synthesis-feasibility-classification}}

Results are additionally classified into a four-tier match system: an exact match (tier 1) or a similar composition (tier 2) are read directly off the returned entries; a partial subsystem match (tier 3) --- where only part of the candidate's chemical system is covered by existing literature --- is where a subsequent CALPHAD-based thermodynamic check (reported in the Stage 3 paper) adds the most value, since the database result alone cannot resolve phase behavior for the untested remainder of the composition; nothing found (tier 4) hands the full composition to that same experimental-plus-modeling cycle with no literature prior to lean on. The tier assigned at this stage is therefore not just a novelty read but a direct input to how much downstream thermodynamic modeling a candidate composition warrants.

\hypertarget{desktop-gui-and-data-sorting-utility}{%
\subsection{Desktop GUI and data-sorting utility}\label{desktop-gui-and-data-sorting-utility}}

The primary researcher-facing interface is a \texttt{customtkinter} desktop application structured around five tabs (new entry, file import, data viewer, quick lookup, summary). All three literature databases are queried upfront on a single ``Calculate \& Preview'' action and cached locally; independent, per-database relevance-cutoff sliders then re-filter the cached results without any further network call, keeping the interactive exploration step fast regardless of how many times the cutoff is adjusted. A separate, narrower command-line tool remains available for fast, scriptable data import when the GUI's interactivity is unnecessary --- a deliberate CLI/GUI split, not legacy duplication: the CLI stays dialogue-only for scripted import from existing files, while anything visual or exploratory goes through the GUI.

\begin{figure}[htbp]
    \centering
    \includegraphics[width=1.0\textwidth]{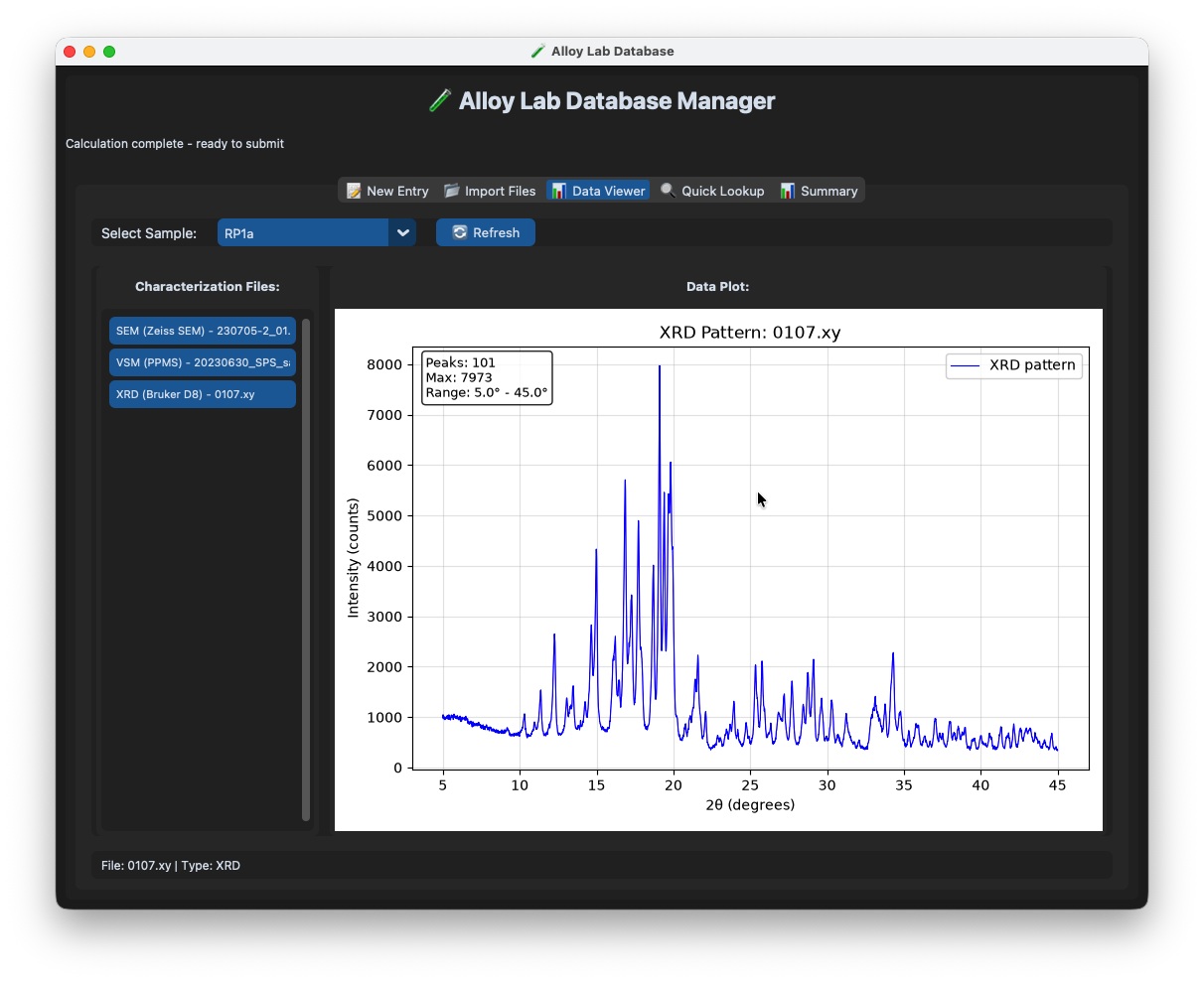}
    \caption{The Data Viewer tab displaying a raw XRD pattern (sample RP1a, Bruker D8): peak count, maximum intensity, and angular range are shown for visual inspection, with no quantitative extraction performed at this stage.}
    \label{fig:xrd-viewer}
\end{figure}

Figure~\ref{fig:xrd-viewer} shows the Data Viewer tab rendering a raw XRD pattern. Consistent with the scope boundary set out in Section~\ref{system-overview}, Stage 1 stops at this visual inspection step: it does not extract peak positions, phase identification, or a lattice parameter from the pattern, since that quantitative extraction is the Stage 2 characterization pipeline.

\begin{figure}[htbp]
    \centering
    \includegraphics[width=1.0\textwidth]{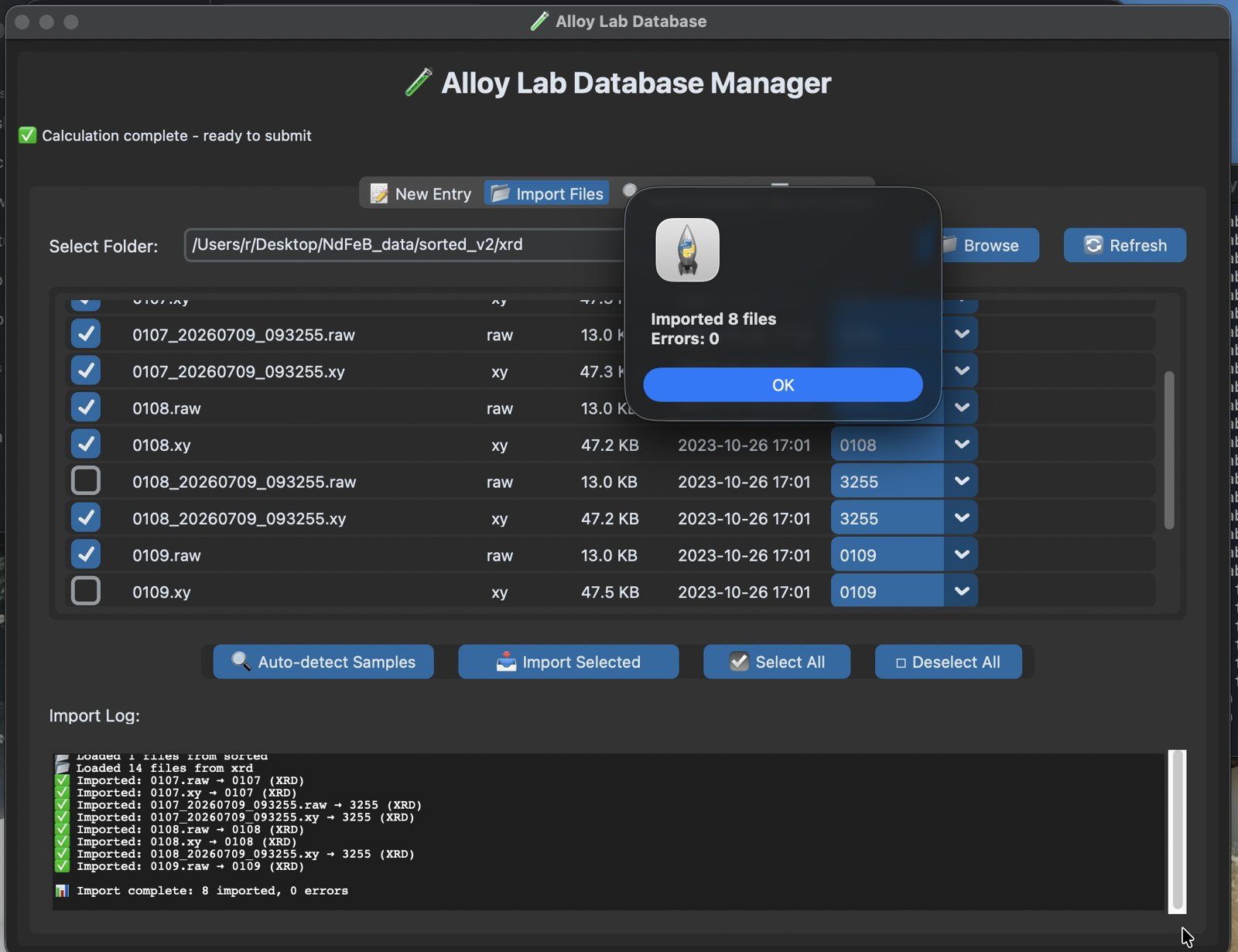}
    \caption{The Import Files tab: raw instrument files auto-detected in a working folder, matched to sample IDs, and imported in a batch, with a running log of what was imported and any errors.}
    \label{fig:import-gui}
\end{figure}

A standalone utility separately auto-classifies files dropped into an unsorted working folder, by both file type (XRD/SEM/VSM/ICP/process notes/images, etc.) and sample ID, using extension- and filename-pattern heuristics --- reducing a recurring manual-triage step common to any lab with heterogeneous instrument output to a single command. Figure~\ref{fig:import-gui} shows the Import Files tab after a batch import: files auto-detected in a working folder, matched to sample IDs, and queued for import rather than sorted and renamed by hand.

\hypertarget{engineering-validation-and-lessons-learned}{%
\section{Engineering Validation and Lessons Learned}\label{engineering-validation-and-lessons-learned}}

We report five engineering findings in detail --- four defects in our own code and one dependency-level rejection --- each chosen because it generalizes beyond this codebase to any small scientific-software project built by a domain expert without formal software-engineering training, a situation we believe is common and underreported.

\hypertarget{silent-parsing-ambiguity}{%
\subsection{Silent parsing ambiguity}\label{silent-parsing-ambiguity}}

\textbf{Defect.} An early version of the composition parser treated a string like ``Cp'' inconsistently: C and P are both real element symbols, so ``Cp'' is a genuine ambiguity (is it one element, Cp --- not a real element --- or two, C and P?), not a malformed input. A permissive parser resolved this silently, with no indication to the user that a judgment call had been made.

\textbf{Fix.} The parser was rewritten to be strict: every element symbol must be immediately followed by an explicit numeric fraction. This converts every ambiguous case into an explicit, user-visible parse error rather than a silent, possibly-wrong interpretation.

\textbf{Generalizable lesson.} A parser's job in scientific software is not just to accept valid input but to actively detect when a domain-specific ambiguity exists; permissive parsing that ``just works'' on the common case is the more dangerous failure mode, because the single output looks exactly as trustworthy as a correct one.

\hypertarget{a-partial-edit-that-broke-a-dependency-chain-invisibly}{%
\subsection{A partial edit that broke a dependency chain invisibly}\label{a-partial-edit-that-broke-a-dependency-chain-invisibly}}

\textbf{Defect.} An edit adding a new distance-cutoff filtering function accidentally deleted an existing helper function's signature while leaving its body behind as orphaned, syntactically valid dead code nested inside the wrong function. This silently broke literature lookups across all three database integrations simultaneously (\texttt{NameError:\ name\ \textquotesingle{}\_is\_better\textquotesingle{}\ is\ not\ defined}), but the defect shipped because the fix was verified by testing only the newly added function in isolation --- not the full call chain a real bug report would exercise.

\textbf{Fix.} The orphaned code was restored as its own properly-scoped function, and the fix was this time verified by exercising the actual, complete call chain (\texttt{MatchResult} $\rightarrow$ \texttt{from\_mp\_results} $\rightarrow$ \texttt{dedup\_by\_formula} $\rightarrow$ \texttt{filter\_by\_distance} $\rightarrow$ \texttt{GUI render}) rather than the single function that had been directly edited.

\textbf{Generalizable lesson.} Verifying a fix against the function that was edited, rather than the full path a bug report actually came from, systematically under-tests exactly the kind of regression most likely in small, single-maintainer codebases: a change whose damage is in what it silently deleted elsewhere, not in what it added.

\hypertarget{a-numeric-type-bug-surfaced-as-a-database-error}{%
\subsection{A numeric-type bug surfaced as a database error}\label{a-numeric-type-bug-surfaced-as-a-database-error}}

\textbf{Defect.} A computed lattice parameter (from the Stage-2 XRD pipeline, built immediately after this one and sharing its persistence layer) was inserted into the database via string formatting, which let a NumPy scalar's internal string representation leak literally into the generated SQL text rather than being converted to a native type first --- producing a database error (\texttt{schema\ "np"\ does\ not\ exist}) with no obvious connection to its numeric-type root cause.

\textbf{Fix.} Explicit casting to a native Python \texttt{float} before binding the value as a parameterized query argument.

\textbf{Generalizable lesson.} Numeric library types (NumPy scalars, pandas values) are not safely interchangeable with native types across a string-based interface boundary; an error message at that boundary (here, a database driver) can point entirely away from the actual root cause, which is worth deliberately checking for whenever a numeric value crosses from a scientific-computing library into a persistence or serialization layer.

\hypertarget{a-credential-hygiene-gap-.gitignore-alone-did-not-close}{%
\subsection{A credential-hygiene gap \texttt{.gitignore} alone did not close}\label{a-credential-hygiene-gap-.gitignore-alone-did-not-close}}

\textbf{Defect.} After migrating the database password out of source code and into a \texttt{.env} file, a commented-out (not deleted) copy of the plaintext password was found still present in a configuration file on a later review --- missed by an initial grep pass that searched only for active, uncommented occurrences. Separately, a data folder containing raw experimental files was added to \texttt{.gitignore} only after it had already been committed; since \texttt{.gitignore} has no retroactive effect on already-tracked files, the folder remained fully present in git history despite appearing absent from the current working tree.

\textbf{Fix.} The commented-out password was deleted and git history was audited directly (\texttt{git\ log\ -\/-all\ -p\ \textbar{}\ grep\ \textless{}string\textgreater{}}) rather than trusting current file contents, to confirm the plaintext value had never actually been committed. The data folder was removed from history with \texttt{git-filter-repo}, verified first on a disposable clone (confirming the path no longer appeared anywhere in history and that repository size dropped accordingly) before the same procedure was applied to the real repository.

\textbf{Generalizable lesson.} Neither a comment-out edit nor a \texttt{.gitignore} entry is itself a security control; the only way to verify whether a secret or sensitive file was ever exposed is to audit git history directly, not current file contents, and a repository should be set private while that audit is pending rather than after.

\hypertarget{a-dependency-rejected-for-silent-corruption-risk}{%
\subsection{A dependency rejected for silent-corruption risk}\label{a-dependency-rejected-for-silent-corruption-risk}}

\textbf{Context.} AFLOW was the fourth materials database considered for the literature cross-referencing pipeline described in Section~\ref{multi-database-literature-cross-referencing}, alongside Materials Project, OQMD, and Alexandria. Its official Python client (version 0.0.11) had not been meaningfully maintained since roughly 2017--2019.

\textbf{Findings.} Three independent problems surfaced during integration, each sufficient on its own to warrant caution: the client's query endpoint returned a live HTTP 404, indicating a moved or retired server-side endpoint rather than a transient outage; the underlying AFLUX query language has an unresolved AND/OR precedence ambiguity in its filter syntax, which the client does not disambiguate or warn about; and, most seriously, the client's keyword-filter objects are implemented as mutable singletons, so a second \texttt{.filter()} call silently mutates and corrupts the filter state left behind by a first call --- meaning two sequential, independently-reasoned queries in the same process can silently contaminate each other's results, with no exception raised and nothing in the returned data to indicate anything went wrong.

\textbf{Decision.} The client was dropped rather than patched or worked around. A patched fork would have meant maintaining a fix for an abandoned upstream project indefinitely; a workaround (a fresh filter object per query, say) would have papered over a defect class --- silent mutation of shared state --- that could plausibly resurface elsewhere in the same dependency. Alexandria was adopted in AFLOW's place and, independently of this decision, turned out to better fill the role of a large, purely computational third database (Section~\ref{multi-database-literature-cross-referencing}).

\textbf{Generalizable lesson.} An unmaintained dependency's risk is not fully captured by whether it currently runs; a confirmed silent-corruption bug in shared, mutable state is a standing risk to every future use of that dependency, not a one-time defect to patch around, and is reason enough to reject it even when a workaround for the immediate symptom is available.

\hypertarget{summary}{%
\subsection{Summary}\label{summary}}

All four code defects share a structure: each produced an output that was plausible-looking rather than obviously broken, and each was caught by exercising the system under real or realistically adverse conditions (a genuinely ambiguous real formula, a real downstream bug report, a real database insert, a real git-history audit) rather than by code review or an isolated unit test. The dependency rejection of Section~\ref{a-dependency-rejected-for-silent-corruption-risk} exhibits the same failure mode one layer removed --- a silent corruption whose output looks as trustworthy as a correct one --- and was caught by the same kind of real-use exercise, in this case an integration attempt rather than a bug report. We take this as the central practical lesson of Stage 1: for a small scientific-software project with no dedicated QA process, validating against real use is doing a substantial share of the work that a larger project's test suite and code review would otherwise do, and that validation has to reach the full path a defect could plausibly take, not just the code that was most recently touched.

\hypertarget{discussion-and-limitations}{%
\section{Discussion and Limitations}\label{discussion-and-limitations}}

\textbf{What this paper deliberately does not report.} Consistent with the software-architecture framing of this work, we do not report specific sample compositions, measured properties, or literature-match outcomes for named samples. Where an example composition is needed to illustrate a design choice, it is representative or synthetic rather than drawn from an actual unpublished sample. This is a reporting choice, not a confidentiality constraint --- the underlying data comes from open, previously published projects --- made because the contribution of this paper is the infrastructure and the engineering process, not the materials-science findings it produces.

\textbf{Single-lab, single-maintainer scope.} The system was built by and for one researcher's own lab. It has not been evaluated across multiple labs, multiple maintainers, or at a scale beyond roughly 50 samples at the time Stage 1 closed. We therefore make no claim that the specific architectural choices here (a single PostgreSQL instance, a desktop GUI as the primary interface, JSONB for composition storage) would scale unchanged to a multi-user or multi-lab deployment; they were right-sized for the problem actually being solved.

\textbf{No formal test suite.} The system has no automated test suite in the conventional sense; validation instead relied on the kind of real-use exercising described in Section~\ref{engineering-validation-and-lessons-learned}. We view this as an honest account of how a time-constrained, single-maintainer scientific-software project actually gets validated in practice, not as a recommended substitute for automated testing where the resources for it exist.

\textbf{Database selection is not exhaustive.} Three materials databases were integrated and a fourth (AFLOW) was evaluated and rejected (Section~\ref{a-dependency-rejected-for-silent-corruption-risk}); this is not a claim that no other public materials database would be worth integrating, only that these three covered the lab's practical needs at acceptable engineering cost.

\textbf{\texttt{literature\_sources} is not yet cross-linked.} The \texttt{literature\_sources} table (Figure~\ref{fig:schema}) is populated with source DOIs from Materials Project's text-mined synthesis records when a sample is saved, but is not yet linked by a foreign key from \texttt{literature\_checks}: Stage 1 only needs the per-check match classification (Section~\ref{four-tier-synthesis-feasibility-classification}), not a durable, queryable link to a specific stored literature record. That link is expected to be added in Stage 3, once the retrieval-augmented synthesis advisor needs to query stored literature entries directly rather than treat each database check as a one-off lookup.

\textbf{Relationship to Stage 2 and Stage 3.} This paper reports Stage 1 in isolation because it reached a stable, validated state before Stage 2 development began, and because each stage's validated scope is large enough to report independently. Stage 2 (validated XRD/VSM/SEM characterization pipelines, an element-fraction table, and a multi-machine server migration) and Stage 3 (a retrieval-augmented literature advisor and a CALPHAD thermodynamic feasibility module) are built directly on top of the schema and conventions established here, and are reported as separate papers.

\textbf{A snapshot, not a live document.} This paper describes the Stage 1 system as a fixed snapshot, frozen once Stage 1 reached the stable, validated state reported here. The underlying codebase is a living project that continues to develop after this report, and the repository is organized precisely so that this snapshot stays accurate: each stage keeps its own subfolder, left untouched once its paper is written, while ongoing work --- including the shared root utilities all stages build on --- continues elsewhere. Where the current system differs from what is described here, the difference should be in a positive direction (a fix, an added capability, broader test coverage) rather than a regression, since that shared foundation keeps improving as later stages are built on top of it. Readers should treat this paper as a dated account of the system at the point Stage 1 closed, not as live documentation of its current state.

\hypertarget{conclusion-and-future-work}{%
\section{Conclusion and Future Work}\label{conclusion-and-future-work}}

We presented Stage 1 of an open-source data infrastructure for alloy synthesis research: a composition calculator, physics-based screening metrics, automated multi-database literature cross-referencing, and a desktop GUI, built end to end by a single domain-expert researcher without formal software-engineering training. We reported the system's architecture and, in detail, five engineering findings --- four defects in our own code (a silent parsing ambiguity, a dependency-breaking partial edit, a numeric-type database bug, and a credential-hygiene gap that \texttt{.gitignore} alone did not close) and one dependency-level rejection (a third-party database client discarded for silent-corruption risk, Section~\ref{a-dependency-rejected-for-silent-corruption-risk}) --- each generalized into a lesson we believe applies broadly to small, similarly-resourced scientific-software projects: that validation has to exercise the full path a defect could plausibly take, and that a config-file comment-out or a \texttt{.gitignore} entry is not itself a security control.

Two follow-on papers extend this work directly. Stage 2 reports validated, instrument-specific characterization pipelines (XRD, VSM, SEM) that attach quantitative measurements to the sample records this system creates, an element-fraction table enabling machine-learning-style queries, and a migration of the database to shared, always-on infrastructure. Stage 3 reports a retrieval-augmented synthesis advisor built on a large materials-science literature corpus and a CALPHAD-based thermodynamic phase-diagram module for pre-synthesis feasibility screening. Together, the three stages are intended to demonstrate that a practicing researcher can build durable, reusable research infrastructure incrementally, validating and reporting each stage before the next begins, rather than deferring all engineering investment until a larger, better-resourced project makes it possible.

\hypertarget{references}{%
\section*{References}\label{references}}

{[}1{]} Guo, S., Ng, C., Lu, J., \& Liu, C. T. (2011). Effect of valence electron concentration on stability of fcc or bcc phase in high-entropy alloys. \emph{Journal of Applied Physics}, 109(10), 103505.

{[}2{]} Yang, X., \& Zhang, Y. (2012). Prediction of high-entropy stabilized solid-solution in multi-component alloys. \emph{Materials Chemistry and Physics}, 132(2--3), 233--238.

{[}3{]} Takeuchi, A., \& Inoue, A. (2005). Classification of bulk metallic glasses by atomic size difference, heat of mixing and period of constituent elements and its application to characterization of the main alloying element. \emph{Materials Transactions}, 46(12), 2817--2829.

{[}4{]} Takeuchi, A., \& Inoue, A. (2000). Calculations of mixing enthalpy and mismatch entropy for ternary amorphous alloys. \emph{Materials Transactions, JIM}, 41(11), 1372--1378.

\end{document}